\documentclass[11pt]{article}

\usepackage{comment}

\usepackage{amssymb}
\usepackage{amsmath}
\usepackage{amsthm}
\usepackage[top=1.5in, bottom=1.5in, left=1.5in, right=1.5in]{geometry}
\usepackage{color}

\newcommand{\condition}{\,\mid\:}
\newcommand{\ppoly}{\rm P/poly}

\newcommand{\pe}{\mbox{\rm P}}

\newcommand{\np}{\mbox{\rm NP}}

\mathchardef\mhyphen="2D

\usepackage{fullpage}
\usepackage[utf8]{inputenc}
\usepackage{amsmath}
\usepackage{setspace}

\usepackage{amssymb}
\usepackage{graphicx}

\usepackage{tikz}
\DeclareRobustCommand{\models}{\mathrel{||}\joinrel\Relbar}

\title{On Lev Gordeev's ``On P Versus NP''}
\author{David E. Narváez\thanks{Supported in part by NSF grant CCF-2030859 to the Computing Research Association for the CIFellows Project.}\\Department of Computer Science\\University of Rochester\\Rochester, NY, USA 14627 \and Patrick Phillips\thanks{Supported in part by NSF grant CCF-2006496.}\\Department of Computer Science\\University of Rochester\\Rochester, NY, USA 14627}

\begin{document}

\maketitle
\begin{abstract}
    In the paper ``On P versus NP,'' Lev Gordeev attempts to extend the method of approximation, which successfully proved exponential lower bounds for monotone circuits, to the case of De Morgan Normal (DMN) circuits. As in Razborov's proof of exponential lower bounds for monotone circuits, Gordeev's work is focused on the \np-complete problem CLIQUE\@. If successful in proving exponential DMN circuit lower bounds for CLIQUE, Gordeev would prove that $\pe\neq\np$. However, we show that Gordeev makes a crucial mistake in Lemma 12. This mistake comes from only approximating operations over positive circuit inputs. Furthermore, we argue that efforts to extend the method of approximation to DMN circuits will need to approximate negated inputs as well.
    
\end{abstract}
\section{Introduction}
Lev Gordeev's recent paper titled ``On P versus NP'' claims to have shown that $\pe\neq\np$~\cite{gordeev2020p}. Gordeev attempts to do so by showing that only an exponential-sized Boolean circuit ($\lor, \land, \neg$) can solve CLIQUE$(k^4, k)$, the problem of checking if a graph of $k^4$ vertices has a clique of size $k$. Since CLIQUE is in $\np$ this would show $\np$ is not a subset of the class of problems solved by polynomial-sized circuits ($\np  \not\subseteq \ppoly$), and since it is commonly known that $\pe \subsetneq\ppoly$, we would have that $\pe\neq\np$.

In 1985, Alexander Razborov proved that monotone circuits ($\lor, \land)$ require exponential size to solve CLIQUE~\cite{razborov1985lower}. Razborov's proof introduced a new technique called the ``method of approximation.'' The method of approximation begins by introducing two new operators $\sqcap$ and $\sqcup$ that approximate $\land$ and $\lor$, respectively. The idea is to define $\sqcap$ and $\sqcup$ so that \textit{no} formula generated from $\sqcap$ and $\sqcup$ is close to solving CLIQUE\@. However, we also want our definitions of $\sqcap$ and $\sqcup$ to be such that at each step where $\sqcap$ or $\sqcup$ is used instead of $\land$ or $\lor$, we introduce at most a relatively small amount of error. If we can define operators $\sqcap$ or $\sqcup$ in this way then we can say that there must have been a lot of total operations in any circuit that actually solves CLIQUE\@.

\begin{figure}[h!]
    \centering
    \begin{tikzpicture}[level/.style={sibling distance=1in/#1}]
    \node {$\leq 7e$}
    child { 
     node {$\leq 3e$}
     child {
      node {$\leq e$} child { node {} } child { node {} }
     }
     child {
      node {$\leq e$} child { node {} } child { node {} }
     }
    }
    child {
     node {$\leq 3e$}
     child {
      node {$\leq e$} child { node {} } child { node {} }
     }
     child {
      node {$\leq e$} child { node {} } child { node {} }
     }
    };
    \end{tikzpicture}
    \caption{Error introduced in monotone circuits from using approximators $\sqcap$ and $\sqcup$ instead of $\land$ and $\lor$.}
    \label{fig:err}
\end{figure}
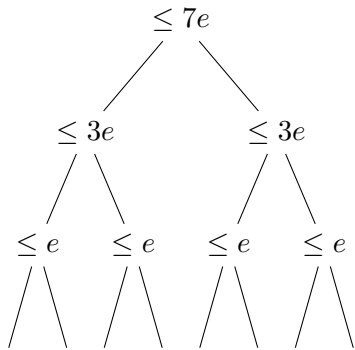

To reiterate the idea a little more carefully, first suppose there is some monotone circuit that correctly solves CLIQUE\@. We then assume the circuit generated by replacing all the $\land$ and $\lor$ operations with $\sqcap$ and $\sqcup$ operations is not close to solving CLIQUE, i.e., there are at least $T$ mistakes over all possible input graphs. However, at each operation in the circuit we also assume that we do not introduce many new mistakes, say at most $e$ new mistakes, by using approximators. Letting $|C|$ be the number of operations in the circuit, we would have that the total number of mistakes is bigger than $T$, but also less than $|C|e$. So we have $|C|e \geq T$ or $|C| \geq \frac{T}{e}$, i.e., these two assumptions would let us say that there are at least $\frac{T}{e}$ total steps in the circuit. A graphical representation is shown in Figure~\ref{fig:err}. 

Gordeev wants to extend the method of approximation to handle negation operators ($\neg$). One key step in handling negation operators is the observation that converting a regular Boolean circuit ($\lor, \land, \neg$) to a De Morgan Normal (DMN) circuit ($\lor, \land$), which allows negated inputs (unlike monotone circuits), requires at most doubling the size of the Boolean circuit. Thus if we can show we need exponential-sized DMN circuits to solve CLIQUE we have shown we need exponential-sized Boolean circuits to solve CLIQUE\@. Because of the similarity between DMN circuits and monotone circuits, one might think that we can largely reuse the method of approximation in the case of DMN circuits. It is a sensible approach: all one has to do is introduce two approximators $\sqcap$ and $\sqcup$, show that at each step they do not add a significant amount of error relative to the exact operators $\land$ and $\lor$, and also show that any formula using $\sqcap$ and $\sqcup$ must have a large total amount of error relative to CLIQUE\@.

Unfortunately, Gordeev's attempt to adapt the method of approximation for the case of DMN circuits contains a mistake on the way to developing the lower bound $T$. Gordeev's approach and mistake are closely related to a similar paper published by Norbert Blum in 2017, which also had the goal of extending the method of approximation to DMN circuits~\cite{blum2017solution}. Before we more precisely discuss the mistake, a better understanding of the method of approximation and an overview of Gordeev's approach and notation are necessary. 

\section{Gordeev's Adaptation of the Method of Approximation}

We should note this summary will by no means be a sufficient replacement for reading Gordeev's paper. That being said, we will try to cover the important steps taken by Gordeev so that the mistake in the approach is evident. First, we begin with some notation that formalizes graph-theoretic notions. The following section is largely taken verbatim from Gordeev. 

\subsection{Graph-Theoretic Background and Notation}
For any $n>0$ and sets $X$, $Y$ we let $%
[ n]=\{ 1,\ldots ,n\} $ and consider products $X\cdot Y=\{\{ x,y\} \condition x\in X\land y\in Y\land x\neq y\}$ and $X^{(2)}=X\cdot X$. We fix $m=k^4$ to be the number of vertices in a graph $G$. Plain (undirected) graphs with $m$ vertices
are nonempty subsets of $[m]^{(2)}$. For any graph 
$\emptyset \neq G\subseteq [ m] ^{(2)}$ we regard
pairs $\{ x,y\} \in G$ as \emph{edges} and define its \emph{vertices} $\mathrm{v}(G)=\{ x\in [ m] \condition (\exists y\in [m]) [\{ x,y\} \in G] \} $. $\mathcal{G}=\mathcal{P}([m] ^{(2)})\setminus\{
\emptyset\} $ will denote the set of all graphs and $\mathcal{K}=\{ K\in \mathcal{G}\condition\| \mathrm{v}(K)\| =k\land K=\mathrm{v}(K) ^{(2)}\}\subsetneq
\mathcal{G}$ the set of \emph{complete graphs} with $k$ vertices.\footnote{Note that Gordeev defines the set of complete graphs with $k$ vertices as $\mathcal{K}=\{ K\in \mathcal{G} \condition \|K\| = k \land K=\mathrm{v}(K)^{(2) }\} $, however we believe this to be a typo and that the intended definition coincides with ours, which replaces $\|K\|$ with $\|\mathrm{v}(K)\|$.} Below we
identify the $\mathrm{CLIQUE}(m,k) $ problem with the set of its affirmative
solutions $G\in \mathcal{G}$ and use abbreviation 
\begin{equation*}
\mathrm{CLIQUE}=\left\{ G\in \mathcal{G} \condition \left( \exists K\in 
\mathcal{K}\right)[K\subseteq G]\right\}.
\end{equation*}

\subsection{Background on Gordeev's Approach}
In the introduction we talked about having ``mistakes,'' i.e., graphs that actually did have a clique of size $k$ but were rejected by our circuit, or graphs that did not have a clique of size $k$ but were accepted by our circuit. It turns out to be more convenient to consider mistakes over only a specific set of ``test graphs'' instead of counting the number of mistakes over all possible graphs. Gordeev defines $\mathrm{POS}=\mathcal{K}$, the set of complete graphs, as a set of positive test graphs that should be accepted. To define NEG, Gordeev first introduces the typical coloring functions, $\mathcal{F}=\{f : [m] \longrightarrow [k-1]\} $, so $\| \mathcal{F}\| =( k-1)^m$. Gordeev then defines $\mathrm{NEG}=\{ C_f \condition f\in \mathcal{F}\}$ where $C_{f}=\{\{ x,y\} \in [m] ^{( 2)
} \condition f(x) \neq f(y)\} \in \mathcal{G}$. Think of $C_f$ as the maximal graph that is properly colored (no two adjacent vertices have the same color) with respect to coloring function $f\in \mathcal{F}$. Clearly, no graph $C_f$ will have a clique of size $k$ since any graph containing a clique of size $k$ cannot be properly colored with $k-1$ colors.

To reiterate, Gordeev is trying to prove that an arbitrary DMN circuit which takes as input a graph $G \in \mathcal{G}$ and then correctly decides if $G \in \mathrm{CLIQUE}$ must have exponential size. Gordeev considers DNFs with both positive and negative literals, which are logically equivalent to arbitrary DMN circuits. Note that if this DNF representation of a DMN circuit has exponential size, it does not guarantee that the DMN circuit has exponential size. 

Each DNF consists of conjunctions of positive literals $\overrightarrow{u}$ and negated literals $\neg\overrightarrow{v}$, where $\overrightarrow{u}=u_{1}\cdots u_{r}$ and $\overrightarrow{v} =v_{1}\cdots v_{s}$, for $r,s\leq \binom{k^{4}}{2}$, are strings of variables. The positive literals represent the edges that must be in $G$ to satisfy the conjunction and the negated literals represent the edges that are not allowed to be in $G$, i.e., the conjunction is satisfied if
$\overrightarrow{u}\subseteq G$ and $\overrightarrow{v}\cap G=\emptyset$. The entire DNF is then viewed as a set $\mathcal{X}$ consisting of pairs $%
\left\langle \overrightarrow{u},\lnot \overrightarrow{v}\right\rangle $. Thus  $G$ is \emph{accepted} by $\mathcal{X}$ if there
exists $\left\langle \overrightarrow{u},\lnot \overrightarrow{v}%
\right\rangle \in \mathcal{X}$ such that $\overrightarrow{u}\subseteq G$ and 
$\overrightarrow{v}\cap G=\emptyset $. Gordeev introduces double graphs as convenient book-keeping to match the tuple $\left\langle \overrightarrow{u},\lnot \overrightarrow{v}\right\rangle$.

\subsection{Double Graphs}
A double graph is a pair of disjoint plain graphs. More precisely, Gordeev defines the set of double graphs as
\begin{equation*}
\mathcal{D}=\{ \langle G_{1},G_{2}\rangle \condition G_{1},G_{2}\in \mathcal{G}\cup \{\emptyset\}\land G_{1}\cap
G_{2}=\emptyset \neq G_{1}\cup G_{2}\}.
\end{equation*}

For any double graph $D=\left\langle D^+,D^-\right\rangle \in \mathcal{D}$ we will call $D^+$ the positive part of the double graph and $D^-$ the negative part of the double graph. Again, the intended use of these double graphs is to view a conjunction of positive and negative literals $\langle \overrightarrow{u}, \neg\overrightarrow{v}\rangle$ as the positive and negative parts of a double graph $D$. A DNF will then be viewed as a set of double graphs $\mathcal{X}\subseteq \mathcal{D}$. Thus with each DNF that we represent as $\mathcal{X}\subseteq \mathcal{D}$, we can relate accepted graphs as follows:
\begin{enumerate}
\item  $\mathcal{X} \models G \Leftrightarrow ( \exists D\in \mathcal{X}) [D\subseteq ^\pm G]$ (read as $\mathcal{X}$ \emph{accepts} $G$),
\item  $\mathrm{ACC}( \mathcal{X}) =\{ G\in \mathcal{G} \condition \mathcal{X}\models G\} $.
\end{enumerate}
Here the notation $D\subseteq^\pm G$ is used to denote that $D^+ \subseteq G$ and $D^- \cap G = \emptyset$. Again, this definition represents the notion that for our DNF to accept a graph $G$ we require the positive edges/literals to be present in $G$ and we forbid the negated edges/literals from being present. Gordeev uses the double graph representation in order to define the approximate operators $\sqcap$ and $\sqcup$, as well as to track accepted and rejected graphs. We intentionally avoid restating Gordeev's definitions of $\sqcap$ and $\sqcup$ because they add unnecessary complication.

\section{Gordeev's Error}

Gordeev's error appears when finding a lower bound on the total number of mistakes for any DMN formula $\varphi$. Gordeev finds different lower bounds for two cases. In case (1) we assume we do not accept any positive test graphs, i.e., $\mathrm{POS}\cap \mathrm{ACC}( \mathrm{APR}( \varphi)) = \emptyset $. In case (2) we assume that we accept at least one graph in POS, i.e., $\mathrm{POS}\cap \mathrm{ACC}( \mathrm{APR}(\varphi))\neq\emptyset $. Here Gordeev uses $\mathrm{APR}(\varphi)$ to denote the set of double graphs we get when evaluating $\varphi$ with approximate operators $\sqcap$ and $\sqcup$, and then uses $\mathrm{ACC}(\mathrm{APR}(\varphi))$ to denote all the graphs accepted by evaluating $\varphi$ with these approximate operators. 

Case (1) is easy: if $\mathrm{POS}\cap \mathrm{ACC}( \mathrm{APR}(\varphi))=\emptyset$ then we made a mistake on every graph in POS so we have our lower bound immediately as $\|\mathrm{POS}\|$. Case (2) is more difficult, and in trying to lower-bound the number of mistakes in case (2) is where Gordeev makes an error. Below are the claim and proof as they appear in Gordeev's paper~\cite{gordeev2020p}.

\noindent
\textbf{Lemma 12 Part 3:}

{\addtolength{\leftskip}{5mm}
$\mathrm{POS}\cap \mathrm{ACC}( \mathrm{APR}( \varphi))\neq \emptyset $ implies $\| \mathrm{NEG}\cap 
\mathrm{ACC}( \mathrm{APR}(\varphi))\|>\frac{1}{3}( k-1) ^m$.
}

\noindent
\textbf{Purported Proof:}

{\addtolength{\leftskip}{5mm}
\noindent
$\mathrm{POS\cap ACC}( \mathrm{APR}( \varphi))
\neq\emptyset $ implies $\mathrm{APR}( \varphi)\neq\emptyset $, so there exists $E\in \mathrm{APR}(\varphi)$, 
$\|\text{\textsc{v}}( E^+)\|\leq\sqrt{k} $. Let $\mathcal{T}=\{ f\in \mathcal{F}\condition ( \forall x, y\in \text{\textsc{v}}(E^+))[x\neq y \implies f(x)\neq
f(y)]\} $. Now $\|\mathcal{T}\|(k-1) ^{-m}=1-\mathbb{P}[ \mathcal{F\setminus T}] >\frac{1}{3}$, i.e., $\| \mathcal{T}\| >\frac{1}{3}( k-1) ^m$
for sufficiently large $m$ (cf\@. proof of Lemma 9), which yields $\|\mathrm{NEG}\cap\mathrm{ACC}(\mathrm{APR}(\varphi))\|>\frac{1}{3}( k-1)^m$.

}
\vspace{1.5mm}\noindent
Here Gordeev correctly asserts that since $\mathrm{POS}\cap \mathrm{ACC}(\mathrm{APR}(\varphi))\neq \emptyset$ we have that $\mathrm{APR}(\varphi)$ contains at least one double graph, call it $E$, where $\|\textsc{v}(E^+)\| \leq \sqrt{k}$ (we know $\|\textsc{v}(E^+)\| \leq \sqrt{k}$ from the definitions of $\sqcap$ and $\sqcup$). Gordeev then defines a set $\mathcal{T}$ that represents all the coloring functions that color no two vertices in $E$ the same, and finds a lower bound for $\|\mathcal{T}\|$ that is correct as well. However, Gordeev goes on to assert that the lower bound of $\|\mathcal{T}\|$ is also a lower bound of $\|\mathrm{NEG}\cap\mathrm{ACC}(\mathrm{APR}(\varphi))\|$. This is an error. 

The problem is Gordeev wants to assert that $\{C_f \condition f\in \mathcal{T}\} \subseteq \mathrm{NEG}$ is contained in $\mathrm{ACC}(\mathrm{APR}(\varphi))$. Recall each $C_f$ is the maximal graph that can be properly colored with respect to coloring function $f$. For each graph $G \in \{C_f \condition f\in \mathcal{T}\}$, we do have the relationship $E^+ \subseteq G$ since $G$ is the maximal graph with respect to coloring function $f$ and $E^+$ is properly colored by each coloring function $f\in \mathcal{T}$ by definition of $\mathcal{T}$. However, the problem is that we cannot say anything about $E^-$, and in particular we cannot say $E^- \cap G = \emptyset$ as would be required for $E$ to accept $G$. Thus the upper bound given for $ \|\mathrm{NEG}\cap\mathrm{ACC}(\mathrm{APR}(\varphi))\|$ is mistaken.

It is easy to construct a concrete counterexample to the third part of Lemma 12 in Gordeev's paper. Consider the formula $\varphi$ with a single conjunction of positive literals that correspond to the edges $[k]^{(2)}$ and negative literals that correspond to all other edges that are $[m]^{(2)}\setminus[k]^{(2)}$. When $\varphi$ is evaluated with exact double graph equivalents of $\land$ and $\lor$ this corresponds to the set $\mathcal{X} \in \mathcal{P}(\mathcal{D})$ that contains the single double graph $E$ where $E^+$ is a complete graph over $k$ vertices and $E^-$ contains all other edges. Formally, we can write this set of double graphs as $\mathcal{X} = \{E\}$ where $E = \langle [k]^{(2)}, [m]^{(2)}\setminus [k]^{(2)}\rangle$.\footnote{Gordeev uses notation such as $\mathcal{G}\setminus k^{(2)}$ to represent $[m]^{(2)}\setminus [k]^{(2)}$, however we think the correct interpretation of $\mathcal{G}\setminus [k]^{(2)}$ is the set of all graphs except for the complete graph of $k$ vertices called $[k]^{(2)}$, whereas $[m]^{(2)}\setminus [k]^{(2)}$ clearly refers to the graph with all edges except for those in $[k]^{(2)}$.} If $\varphi$ is evaluated with approximate operators $\sqcap$ and $\sqcup$, then $E^-$ remains unchanged and $E^+$ is replaced by some subgraph of $[k]^{(2)}$  with less than $\sqrt{k}$ vertices.\footnote{See Gordeev's definitions of $\sqcap$ and $\sqcup$ for justification.} Formally, we can write this as $APR(\varphi) = \{E^{\sqrt{k}}\}$ where $E^{\sqrt{k}} = \langle E^{\sqrt{k}^+}, [m]^{(2)}\setminus [k]^{(2)}\rangle$ and $E^{\sqrt{k}^+} \subsetneq [k]^{(2)}$ such that $\mathrm{v}(E^{\sqrt{k}^+}) \leq \sqrt{k}$. Then it is easy to see that we have $\mathrm{POS}\cap \mathrm{ACC}( \mathrm{APR}( \varphi)) =\{[k]^{(2)}\}\neq\emptyset$ since there is exactly one clique that contains a subgraph of $[k]^{(2)}$ and no other edges, namely $[k]^{(2)}$. But we also would have that $\| \mathrm{NEG}\cap 
\mathrm{ACC}(\mathrm{APR}(\varphi))\|\leq \|\mathrm{ACC}(\mathrm{APR}(\varphi)) \| \leq 2^{\|[k]^{(2)}\|}<\frac{1}{3}(k-1)^{m}$ for $m\gg0$, which Gordeev assumes.

This shows that Gordeev's lower bound of $\frac{1}{3}(k-1)^{m}$ is mistaken. Furthermore, if we do not constrain the negative variables, then Gordeev's approximators will not be able to get a lower bound better than $2^{\|[k]^{(2)}\|}$ for the total number of mistakes made. This bound is insufficient, and prevents us from using the method of approximation. The issue here is more of a systematic problem caused by the fact that operations over the negated variables are never approximated. 

Gordeev intentionally avoids approximating the negative variables, likely because exactly evaluating negative variables allows us to easily find strong bounds on the error $e$ introduced by each replacement of $\sqcap$  (respectively $\sqcup$) for $\land$ (respectively $\lor$). This is probably the reason Blum~\cite{blum2017solution} similarly avoided approximating negative variables. Blum was then also met with difficulty in trying to find a minimum total error $T$. In a somewhat similar manner to Gordeev, Blum assumed one could say things about graphs that are accepted by approximating an arbitrary formula $\varphi$ that would only hold in monotone circuits. The above counterexample and other failed efforts provide strong evidence that finding a minimum total error $T$ for these circuit approximations will require an approximation over the negative variables as well.

\section{Conclusion}
Gordeev's mistake leads to the subsequent, and arguably most central, theorem of the paper to fail. Without a lower bound on the total number of mistakes $T$ for any circuit using the approximate operators $\sqcap$ and $\sqcup$, the entire strategy does not get off the ground. 

Overall, it does not seem obvious that an adaptation of the method of approximation is doomed to fail in proving exponential circuit lower bounds. However, it appears that using the method of approximation while exactly evaluating operations over negative literals will not be met with much success. We think a more promising direction is to investigate approximation schemes over both positive and negative variables that still allow us to find adequate bounds on the error introduced by each step of approximation.

\section*{Acknowledgments}

Thanks to Michael C. Chavrimootoo, Lane A. Hemaspaandra, and Melissa Welsh for their helpful comments on earlier versions of this paper. Any remaining errors are the responsibility of the authors.

\bibliography{citations.bib}

\end{document}